# Prospects of carbyne applications in microelectronics

Enormous interest to graphene and nanotubes over the past years have been reasoned by an opportunity to make carbon transistors the fastest in the transistor pool leaving behind even InSb ones. In conventional semiconductors electron mobility is proportional to temperature as the lower is temperature, the weaker are lattice vibrations (so-called phonons), on which electrons scatter, and the higher is mobility. Electron mobility in grapheme displays an unusual behavior – it reaches the maximum at room temperature (of importance is the right choice of the substrate). Lattice vibrations in grapheme are so weak that secondary effects such as admixes or a substrate play here a more essential role. Graphene transistors have a monoatomic layer of pure graphite in the form of a tape. Such tape can be prepared using electron beam lithography by cutting a strip from a graphene sheet, yet here difficulties with the integration into the existing microelectronic technology emerge. And even more challenging is mass production of nanotube-based nanotransistors.

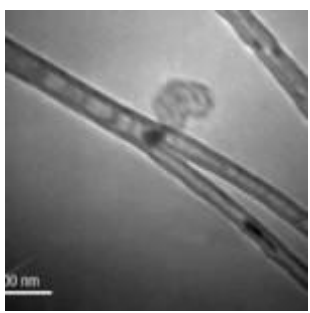

Fig.1. A transistor based on one Y-shaped nanotubes.

Today a lot of individual operating transistors both based on grapheme and on nanotubes have been fabricated, however neither of them is ready for mass production. Provided that the "carbon transistor" is made integrable into the existing silicon technology, whose potential is believed to be shrinking, the sector may see revolutionary changes.

The technology of oriented carbyne film fabrication [1], unlike other carbon forms, can be readily tailored to the existing production processes. In the course of carbyne growing on the substrate kind of a hexagonally packed "forest" of carbon sp1-macromolecules oriented precisely perpendicular to the surface is generated.

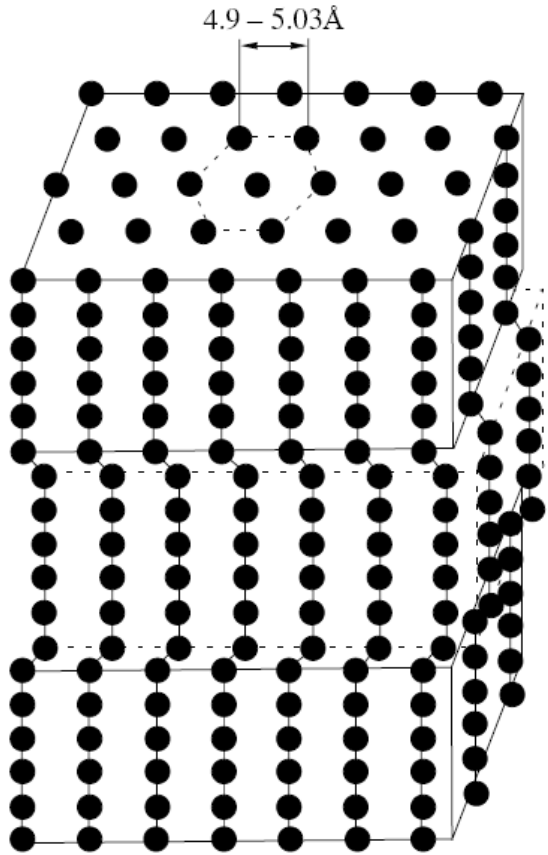

Fig.2. Carbyne structure.

We prepared carbyne transistors on standard 0.5-micron equipment and managed to prove excellent recurrence of the measured characteristics.

As distinct from the conventional current control schemes in carbon transistors, we controlled current through a dielectric that emerged due to a unique capability of carbyne to inject carriers through any energy barrier (Fig.3) [2].

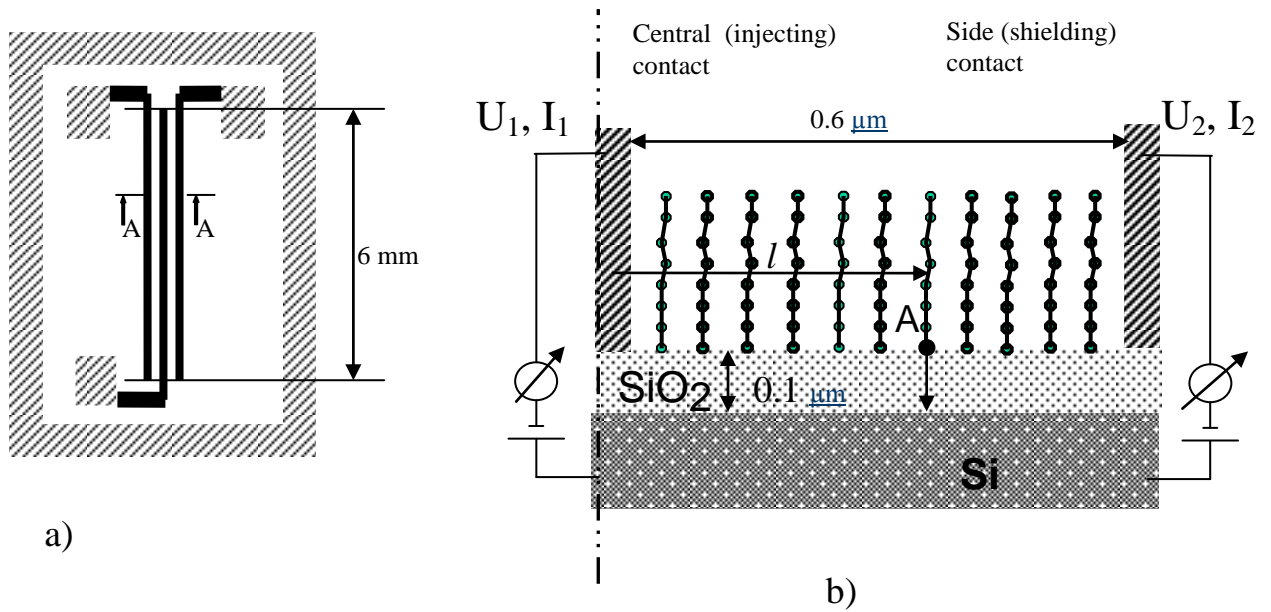

Fig. 3. Specimens and the experiment scheme.

a) the TiN-contact microstructure (black and shaded areas). Top view.

b) the A-A half-section and measurement scheme. Point A denotes the position of the energy barrier, through which charge carriers are injected to $SiO_2$; their pathway is shown by arrows.

The impact of the side electrodes potential on current through $SiO_2$ was found out and a circuit of the specimen inclusion in the field-effect transistor mode was designed. Fig. 4 shows the plots that demonstrate field effect in a "thick" 750 A film.

The experiment scheme details: the volt-ampere characteristics (VAC) was measured with the off-line neighbor electrode (without shielding) and with the side electrodes kept at the Source potential (with shielding). The choice of injection control of holes rather than of electrons was reasoned by the former being nearly an order of magnitude more efficient, which is a factor that reduces parasite current of the Gate. The stepwise switchover of the functional VAC can be assumingly explained by the fact that when current achieves a certain value the field across the chains stops being controlled by the neighbor electrode and falls under control of the space charge in silicon oxide. The latte can be supported by the VAC visible rectification after the current "upset" in double logarithmic coordinates with slope 2, which is consistent with the quadratic law for space-charge-limited (SCL) currents.

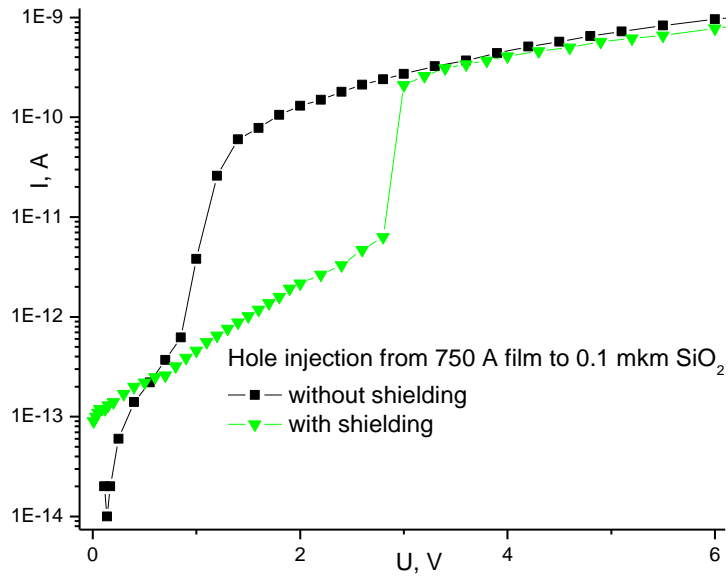

a) Rectification exp(U), which corresponds to the thermal injection of carriers. The difference in the characteristics is conditioned by field effect from shielding side electrodes.

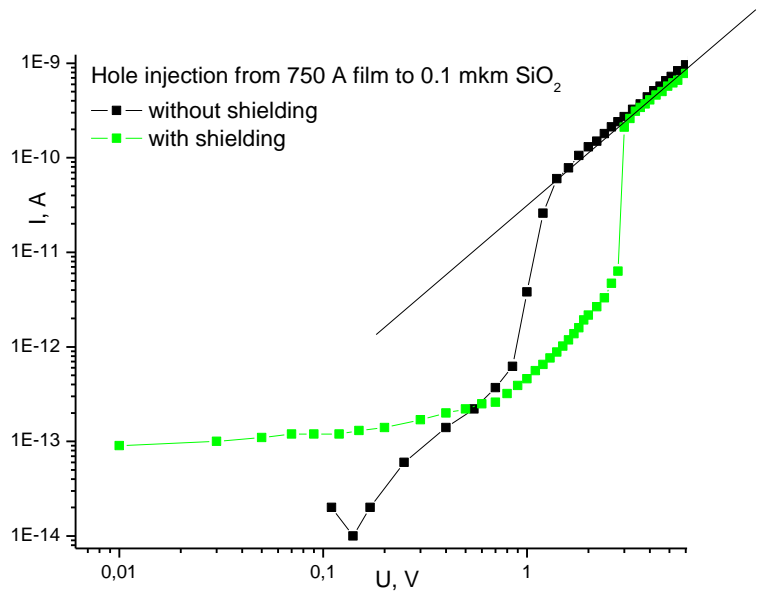

б) Rectification $U^{\alpha}$, values $\alpha= 2$ can be viewed, which is consistent with quadratic SCL current law.

Fig. 4. VACs demonstrating field effect from side electrodes in different rectifying coordinates.

Fig. 5 shows the transfer characteristic of the field-effect transistor on a thin 30 A carbyne film. Current through $SiO_2$ is controlled by the potential of side electrodes. To diminish parasite impact of non-insulated Gate current, control of hole injection current was implemented by the Gate included to the electron injection mode. Hole current shown in Fig. 6, as compared to electron current with off-line neighbor electrodes, indicates how large is the difference in the efficiency of their injections.

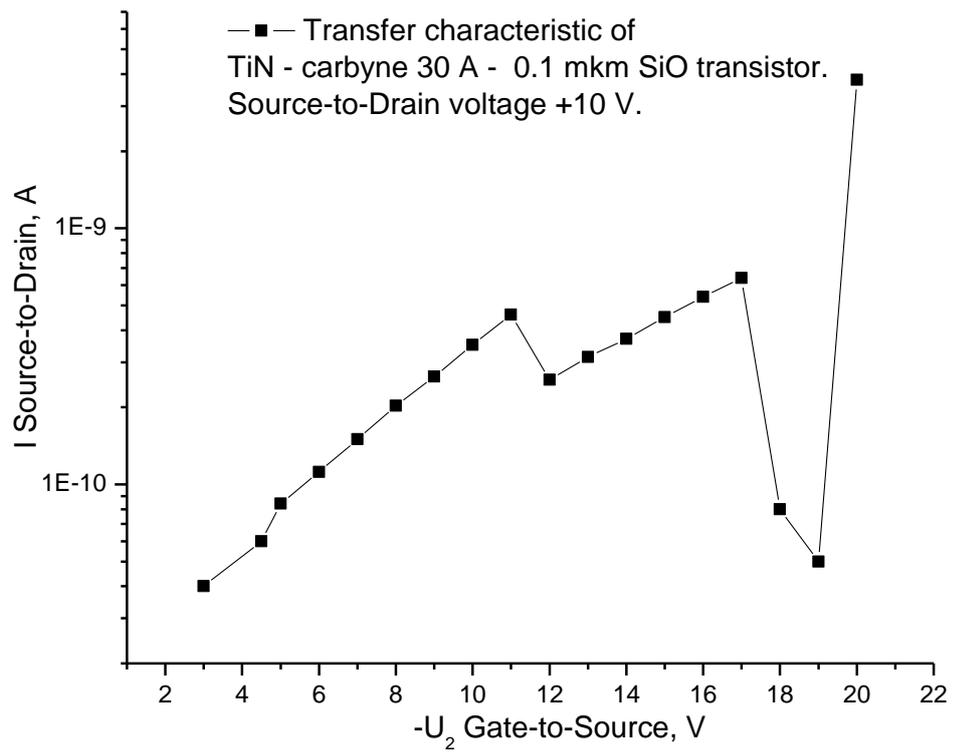

Fig. 5. Transfer characteristic of a carbyne-based field-effect transistor.

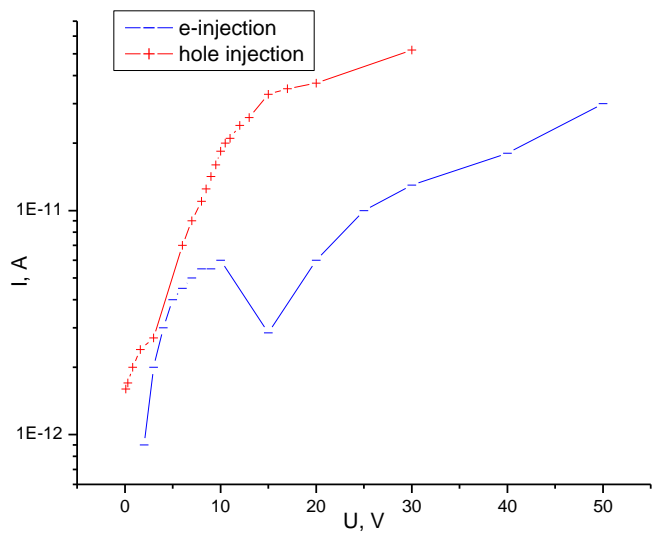

Fig. 6. Hole and electron Source-to-Drain currents with the off-line Gate.

The carrier injection experiments (not only of electrons but also of holes) to the dielectric demonstrated long time of the electrode's life even if very thin films and large current densities are employed. Thin films have a better injecting ability than thick ones as well as a more pronounced field effect, which is likely to indicate that the practically meaningful carbyne phase is obtained in immediate proximity to the substrate surface. This effect had been registered in earlier experiments as an increase in the amorphization degree and defined as an "organizing property of the surface".

Of special note is the specimen forming phenomenon observed in the experiments. Where maximal current values had been achieved from the central electrode to the substrate the modification of the characteristics occurred in the transistors: VAC became unsymmetrical resembling VAC of a diode. Assumingly, the $SiO_2$ layer modification proceeded by the breakdown scenario. The carbyne layer functions here as a breakdown catalyst: the carbyne layer-free specimens prepared by ion etching withstood voltage up to 100 V that corresponds to the electric field of $10^6$ V/cm. Hence carbyne facilitates greatly control of the $SiO_2$ dielectric molding. According to [3] the latter effect can be reversible and be regarded as determinant for furthering design of nanoswitches and memory.

The described transistor can be scaled up in a rather broad range – starting from that prepared by us up to the monomolecular one because the key mechanism here is the inter-chain charge transfer. It currently seems impossible to create a transistor on one sp1-macromolecule within the existing standard technology, however, as the miniaturization evolves, this will become quite feasible.